\documentclass[final]{aipproc}
\layoutstyle{6x9}
\usepackage{graphicx}
\usepackage{epsf}
\usepackage{wrapfig}
\usepackage{sublabel}
\usepackage{epsfig}
\usepackage{amsmath}

\newcommand{\beq}{\begin{eqnarray}}
\newcommand{\eeq}{\end{eqnarray}}

\newcommand{\be}{\begin{equation}}
\newcommand{\ee}{\end{equation}}

\def\b0{{\mbox{\boldmath$0$}}}

\def\b0{{\mbox{\boldmath$0$}}}

\def \b #1{ {\bf #1}}

\def \b #1{ {\bf #1}}

 \newcommand\beqn{\begin{eqnarray}}
 \newcommand\eeqn{\end{eqnarray}}

\def\beqy{\begin{eqnarray}}
\def\eeqy{\end{eqnarray}}

\def\Re{\,\mbox{Re}\,}

\def\sqelha{\sigma_{qel}^{hA}}
\def\sthn{\sigma_{tot}^{hN}}

\def\stota{\sigma_{tot}^{hA}}
\def\sela{\sigma_{el}^{hA}}

\def\sinha{\sigma_{in}^{hA}}

\def\eeq{\end{equation}}

\def\beqy{\begin{eqnarray}}
\def\eeqy{\end{eqnarray}}

\newcommand{\ber}{\begin{displaymath}}
\newcommand{\eer}{\end{displaymath}}
\newcommand{\bey}{\begin{eqnarray}}
\newcommand{\eey}{\end{eqnarray}}

\def\beqy{\begin{eqnarray}}
\def\eeqy{\end{eqnarray}}

\begin{document}
\vskip -2mm
\date{\today}
\vskip -2mm
\title{Nucleon-Nucleon Correlations and Gribov Inelastic Shadowing in
High Energy Collisions}
\author{C. Ciofi degli Atti}{
  address={Department of Physics, University of Perugia and
    Istituto Nazionale di Fisica Nucleare, Sezione di Perugia,
    Via A. Pascoli, I-06123, Italy}=thanks{Work in collaboration with UTFSM, Val  paraiso,
    CHILE}}
\begin{abstract}
The  effects of nucleon-nucleon (NN) short range correlations (SRC) and
Gribov inelastic shadowing (IS) in high energy collisions are
illustrated.
\end{abstract}
\maketitle

The nuclear quantity which enters most Glauber-like calculations is
  the modulus squared of the nuclear wave function $\psi_0$, whose exact
expansion
 \cite{glauber}
 is usually approximated by the lowest order, fully
uncorrelated term, {\it viz} \beqn \left|\,\psi_0({\bf
r}_1,...,{\bf r}_A)\,\right|^2=\prod_{j=1}^A\,\rho_1({\bf r}_j)
\,+\,\sum_{i<j}\,\Delta({\bf r}_i,{\bf
r}_j)\hspace{-0.1cm}\prod_{k\neq i,j}\rho_1({\bf
r}_k)\,+.....\simeq \prod_{j=1}^A\,\rho_1({\bf r}_j).
\label{psiquadro}
 \eeqn
Here the \textit{two-body contraction} ${\Delta({\bf r}_i,{\bf
r}_j)}\,=\,\rho_2({\bf r}_i, {\bf r}_j)\,-\,\rho_{1}({\bf
r}_i)\,\rho_{1}({\bf r}_j)$, contains the effect of SRC,
represented by a hole in  the two-body density  $\rho_2$ at short
\begin{figure}
\centerline{\includegraphics[width=9.5cm,height=7.0cm]{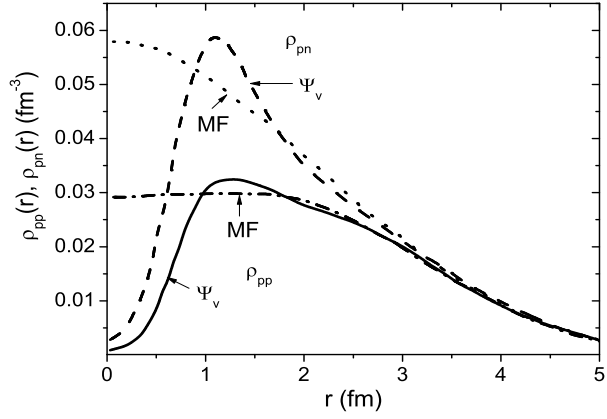}
 \vspace{-0.5cm}\caption{The proton-neutron $(pn)$ and proton-proton
$(pp)$ two-body densities in $^{16}O$ calculated \cite{panda}
within a mean field model (MF)and by solving the many-body problem
with a realistic NN interaction ($\Psi_v$). The behavior of
$\rho_{NN}$ at relative distances $r \leq 1.5\div 2\,\, fm$ is
governed by the repulsive short-range core and by the attractive
intermediate-range tensor force (after Ref. \cite{panda}).}}
  \label{Fig1}
\end{figure}
inter-nucleon separations (see Fig. \ref{Fig1}). SRC   lead to an
additional contribution to the  nuclear thickness function as
follows
  (${\bf r}_i=\{{\bf s}_i,z_i\}$) \cite{totalnA}
\be
  \Delta T_A^h(b)=
  \frac{1}{\sthn}
  \int d^2{\bf s}_1\,d^2{\bf s}_2\,
  \Delta^\perp_A({\bf s}_1,{\bf s}_2)
  \Re \Gamma^{pN}({\bf b}-{\bf s}_1)\,
  \Re \Gamma^{pN}({\bf b}-{\bf s}_2),
  \label{55}
  \eeq
\noindent where $\Delta^\perp_A({\bf s}_1,{\bf s}_2)$
  is the transverse two-nucleon contraction and the total thickness function
is ${\widetilde T}_A^h= T_A^h- \Delta T_A^h$. The thickness
functions of $^{12}C$ and $^{208}Pb$ at HERA-B energies are shown
in Fig. \ref{Fig2}. SRC increase the thickness functions and,
consequently, the total neutron-nucleus cross section at high
energies, making the nucleus more opaque \cite{totalnA},
 unlike  Gribov IS corrections
  which increase nuclear transparency.
An exhaustive calculation  \cite{nashboris} of the total, $\stota$, elastic,
$\sela$, quasi-elastic, $\sqelha$, inelastic, $\sinha$, and
diffractive dissociation hadron-nucleus cross sections, which include both SRC and Gribov
IS summed to all orders by the dipole approach
\cite{KLZ,boris1,boris2},  confirms the opposite roles played by SRC
and IS, with the total contribution to the thickness
function due to SRC and Gribov IS reading as follows
\begin{table}[!htp]
 \begin{tabular*}{0.78\textwidth}{@{\extracolsep{\fill}}c| c c c c }\hline\hline
$^{208}Pb$& Glauber & Glauber & q-2q model & 3q model \\
& & +SRC & +SRC & +SRC \\\hline $\sigma_{tot}^{NA}\,\,\,  [mb]$ &
3850.63 & 3885.77 & 3833.26 & 3839.26 \\\hline $\sigma_{el}^{NA}
\,\,\,[mb]$  & 1664.76 & 1690.48 & 1655.70 & 1660.67 \\\hline
$\sigma_{sd}^{NA}\,\,\, [mb]$ & - & - & 2.62 &    0.59 \\\hline
$\sigma_{sd+g}^{NA} [mb]$& - & - & 2.58 & 2.56
\\\hline $\sigma_{qe}^{NA}\,\, \, [mb]$  &  120.92 &  112.65 & 113.37 & 113.88
\\\hline $\sigma_{qsd}^{NA}\,\,\, [mb]$ & -       & -       & -2.08 & -2.62
\\\hline $\sigma_{tsd}^{NA}\,\,\, [mb]$ & -       & -       & 17.55 & 17.63
\\\hline $\sigma_{dd}^{NA}\,\, \, [mb]$  & -       & -       & -2.08 & -2.62
\\\hline\hline
\end{tabular*}
 \caption{Various $p-^{208}Pb$ cross sections at LHC energies (after Ref. \cite{nashboris}).}
\end{table}
 \begin{eqnarray}
  &&\hspace{-0.5cm}\Delta T_A^{dip}( b,{\bf r}_T,\alpha)=\nonumber\\
  &&\hspace{-0.5cm}=\frac{1}{\sigma_{dip}(r_T)}
  \int d^2{\bf s}_1\,d^2{\bf s}_2\,
  \Delta^\perp_A({\bf s}_1,{\bf s}_2)
  \Re \Gamma^{{\bar q} q,N}({\bf b}-{\bf s}_1,{\bf r}_T,\alpha)
  \Re \Gamma^{{\bar q} q,N}({\bf b}-{\bf s}_2,{\bf r}_T,\alpha).
  \label{500}
  \end{eqnarray}
\begin{figure}
\vspace{-0.5cm}
\centerline{\includegraphics[width=8.0cm,height=7.0cm]{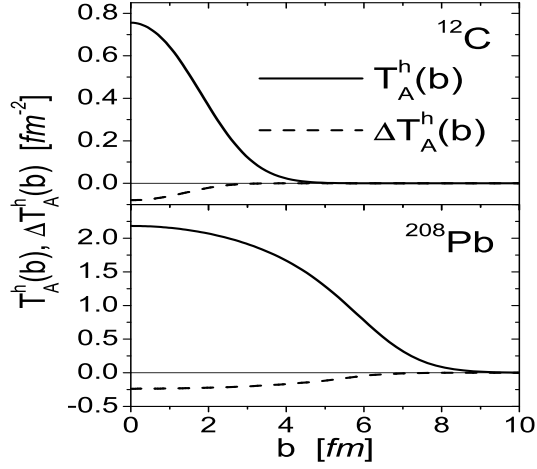}
\caption{The thickness function $T_A^h(b)$ and the correlation
contribution ($\Delta T_A^h(b)$)   in $p-^{12}C$ and $p-^{208}Pb$
collisions  at HERA-B energies. The total thickness function is
given by ${\widetilde T}_A^h= T_A^h- \Delta T_A^h$ (after Ref.
\cite{nashboris}).}}
  \label{Fig2}
  \vspace{-0.5cm}
\end{figure}
SRC and Gribov IS affect also the number of inelastic
collisions $N_{coll}^{hA}=A\,\sigma_{in}^{hN}/\sigma_{in}^{hA}$  which is the normalization
factor  used to obtain the nucleus to
nucleon ratio of the cross section of a hard reaction. The results
of  calculations of Refs. \cite{nashboris,ennecoll}, performed with  realistic one- and two-body
densities and correlation functions from Ref. \cite{ACMprl}, are shown in Tables 1 and 2.
\begin{table}[!h]
\begin{tabular*} {0.98 \textwidth}{@{\extracolsep{\fill}}c| c
c c c c c c}\hline\hline & & & GLAUBER & & & \\ \hline &
$\sigma_{in}^{NN}\:[mb]$ & $\sigma_{tot}^{NA}\:[mb]$ &
$\sigma_{el}^{NA}\:[mb]$ & $\sigma_{qel}^{NA}
\:[mb]$ & $\sigma_{in}^{NA}\:[mb]$& $N_{coll}$\\
\hline RHIC &42.10  &3297.56 &1368.36 &66.06 &1863.14 &4.70 \\
\hline LHC  &68.30  & 3850.63 &1664.76 &120.92 &2064.95 &6.88\\
\hline \hline & & & GLAUBER+SRC & & & \\ \hline &
$\sigma_{in}^{NN}\:[mb]$ & $\sigma_{tot}^{NA}\:[mb]$ &
$\sigma_{el}^{NA}\:[mb]$ &
$\sigma_{qel}^{NA}\:[mb]$ & $\sigma_{in}^{NA}\:[mb]$  & $N_{coll}$\\
\hline RHIC &42.10  &3337.57 &1398.08 &58.47 &1881.02 &4.65\\
\hline LHC  &68.30  & 3885.77& 1690.48&112.65 & 2082.64&6.82\\
\hline \hline & & & GLAUBER+SRC+GRIBOV($q-2q$)& & & \\ \hline
& $\sigma_{in}^{NN}\:[mb]$ & $\sigma_{tot}^{NA}\:[mb]$ & $\sigma_{el}^{NA}\:[mb]$ & $\sigma_{qel}^{NA}\:[mb]$ & $\sigma_{in}^{NA}\:[mb]$  & $N_{coll}$\\
\hline RHIC &42.10  &3228.11 &1314.04 &71.99 & 1842.08 &4.75 \\
\hline LHC  &68.30  &3833.26 &1655.70 &113.37 & 2064.19 &6.88 \\
\hline \hline
 \end{tabular*}
 \caption{Number of inelastic collisions $N_{coll}$  in $p-^{208}Pb$ scattering at
  RHIC and LHC energies (after Ref. \cite{ennecoll}).}
\end{table}
The behavior of $N_{coll}^{NA}$ is entirely governed by the
non-diffractive $\sigma_{in}^{NA}$ which, as shown in Table 2, is
decreased by SRC and increased by Gribov IS. The effects of both
SRC and Gribov IS amount to few percent in agreement with the
results of the calculation of deuteron-gold scattering
\cite{boris1}. Concerning  the high energy collision of two
nuclei, $A$ and $B$,  the correlation contribution to the
thickness function can be written as follows \beqn
&&{\Delta{T}}^h_{AB}(b)= \frac{1}{\sigma^{NN}_{tot}}\,A_A\,A_B^2 \times\nonumber\\
&&\times\int d^2\,s_{A} \rho _A({\bf s}_A) \int d^2s_{B1}
d^2s_{B2} \Delta^{\perp}_B({\bf s}_{B1},{\bf s}_{B2})
\Gamma^{NN}({\bf  b} - {\bf s}_A + {\bf s}_{B1})\Gamma^{NN}({\bf b}
- {\bf
s}_A + {\bf s}_{B2}) +\nonumber\\
&&+\{ A\longleftrightarrow B\}
 \label{Nucleus_fin2}
 \eeqn
 where the 1st term  describes the interaction of a nucleon in $A$ with
 two correlated nucleons in $B$ and the 2nd term in figure brackets viceversa.
The thickness function including the effects of SRC in
$^{208}Pb-^{208}Pb$ scattering at RHIC energies is shown in Fig.
\ref{Fig3}; it can be seen that SRC can appreciably affect  the usual definition of
the number of collisions
$N_{coll}^{AB}(b)$.
\section{Acknowledgment}
A stimulating collaboration with Boris Kopeliovich, Irina
Potashnikova and Ivan Schmidt, Departamento de F\'isica,
Universidad Federico Santa Mar\'ia, Valpara\'iso, Chile, is
gratefully acknowledged.
\begin{figure}
\vspace{-0.5 cm}
\centerline{\includegraphics[width=9.5cm,height=7.0cm]{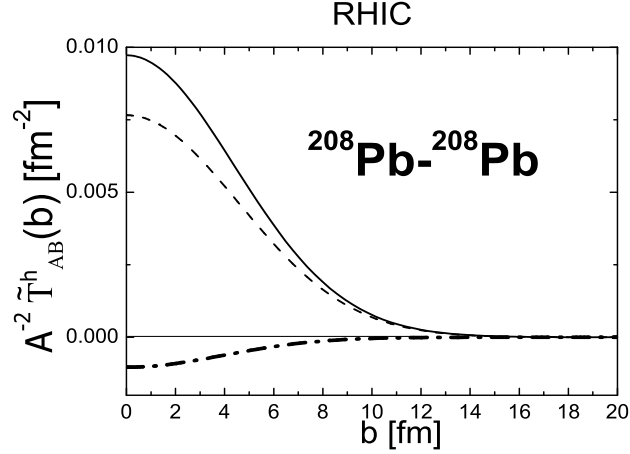}
\vspace{-0.3cm} \caption{Uncorrelated thickness function
$T_{AB}^h(b)/A^2$ (dash);  correlation contribution $\Delta
T_{AB}^h(b)/A^2$  (dot-dash);  total thickness function
  $ {\widetilde T}_{AB}^h/A^2 = [T_{AB}^h(b) -2 \Delta T_{AB}^h(b)]/A^2$ (full)
   in $^{208}Pb
-^{208}Pb$ collisions at RHIC energies  (after Ref.
\cite{ennecoll}).}}
  \label{Fig3}
\end{figure}
\bibliographystyle{aipproc}

\end{document}